\documentclass[twocolumn,showkeys]{revtex4-1}

\usepackage{graphicx,amsmath,color}

\begin{document}

\title{Topological gap solitons in equidistant lithium niobate waveguide arrays}

\author{Andrey V. Gorbach}

\affiliation{Centre for Photonics and Photonic Materials, Department of Physics, University of Bath, Bath BA2 7AY, United Kingdom}
\email{Corresponding author: A.Gorbach@bath.ac.uk}

\begin{abstract}
Equidistant 1D arrays of thin film lithium niobate waveguides can exhibit non-trivial topology due to a specific interplay between inter- and intra-modal couplings of two families of guided modes \cite{Beer2023}. In this work we analyze two-colour spatial solitons, emerging due to $\chi_2$ nonlinear interactions between the modes of non-trivial topology in the fundamental harmonic field, and modes of trivial topology in the second harmonic field. We discuss solitons localized in the bulk of the array (bulk solitons), and at an edge of a finite-size array (edge solitons). The latter emerge due to the nonlinear interactions between a topological edge mode in the fundamental harmonic and bulk modes in the second harmonic. We reveal that for each type of soliton, bulk or edge, there generally exist two families of solutions with different internal structures and ranges of propagation constants. All bulk solitons can only be excited above a certain power threshold dictated by the coupling strength in the second harmonic field and the phase matching between the fundamental and second harmonics. The power threshold for edge solitons generally appears to be much lower, and, by tuning the phase matching, it can be reduced to zero.
\end{abstract}


\pacs{x}

\maketitle

\section{Introduction}
Nonlinear topological photonics is a rapidly developing research field that merges the principles of nonlinear optics and topological phases \cite{Smirnova2020, Szameit2024}. Many novel wave phenomena emerge as the result of the interplay between nonlinearity and topology, including topological solitons \cite{Skryabin2016,Solnyshkov2017,Smirnova2019,Guo2020,Mukherjee2021,Kartashov2022,Ivanov2022,Pernet2022}, frequency conversion and generation of topologically protected photon pairs \cite{Kruk2018,Mittal2018,Lan2020}, topological parametric oscillation \cite{Roy2022,Jiang2024}, topological lasing \cite{Bandres2018,Dong2024}. Inspired by the well-known analogy between tight-binding models in condensed matter physics and the respective models in photonics \cite{Photonic_Crystals_textbook}, most of the studies so far have been focused on periodic arrangements of weakly coupled resonators or waveguides, which mimic the structures of crystals. Examples are the photonic analogues of the so-called Su–Schrieffer–Heeger chain \cite{Malkova2009, Blanco-Redondo2016} and 'photonic graphene' \cite{Plotnik2013}. In such systems, different frequency components of light, interacting via $\chi_3$ \cite{Kruk2018,Guo2020,Pernet2022} or via $\chi_2$ nonlinearities \cite{Lan2020,Kartashov2022,Ivanov2022}, experience the same lattice topology imposed by the fixed structure.

More recently, a new methodology of creating topological photonic structures has emerged, whereby interactions between different families of modes of the individual photonic elements is exploited to create topologically non-trivial phases \cite{Caceres-Aravena2020, Savelev2020, Jiang2022, Beer2023, Mazanov2024, Liu2024}. Notably, the resulting structures often appear to have much simpler geometries, compared e.g. to the above Su–Schrieffer–Heeger-like or graphene-like photonic structures. In particular, the emergence of non-trivial topology and associated topological edge states was recently reported in equidistant arrays of identical multi-mode waveguides \cite{Beer2023}. Such a reduction in the geometrical complexity is helpful for the development of compact and easy-to-manufacture topological photonic devices. Even more importantly, this new class of topological structures represents an interesting platform for studies of novel nonlinear topological wave phenomena. Unlike in the conventional topologic photonic structures, here different interacting frequency components will not necessarily experience the same lattice topology. This is because the topology is created by an intricate interplay between various inter- and intra-modal couplings in this case, and such couplings will strongly vary with wavelength. Thus, in our recent work we demonstrated that $\chi_2$-induced interactions between topological edge states at the fundamental frequency and bulk states of trivial topology in the second harmonic can lead to interesting regimes of topological gain in 1D arrays of multi-mode lithium niobate waveguides \cite{Gorbach2024}. It was also shown that, under certain conditions, such interactions can lead to the emergence of nonlinear two-colour topological edge states. 

In this work, we perform detailed studies of nonlinear edge- and bulk localized states (topological gap solitons) in lithium niobate waveguide arrays. Such nonlinear states represent three-component solitons, whereby two types of guided modes at the fundamental frequency interact with the second harmonic field via $\chi_2$ nonlinearity, leading to the localization in all three components either in the bulk of the array, or at an edge of a finite-size array. The propagation constant of the fundamental frequency components of such solitons resides in the topological gap, and the propagation constant of the second harmonic component is in the semi-infinite gap. We identify parameter regimes of existence of bulk and edge solitons, discuss power thresholds associated with different types of solitons, and how these thresholds can be varied by adjusting the phase-matching condition between the fundamental and second harmonics.

\section{The model}

\begin{figure}
\includegraphics[width=0.45\textwidth]{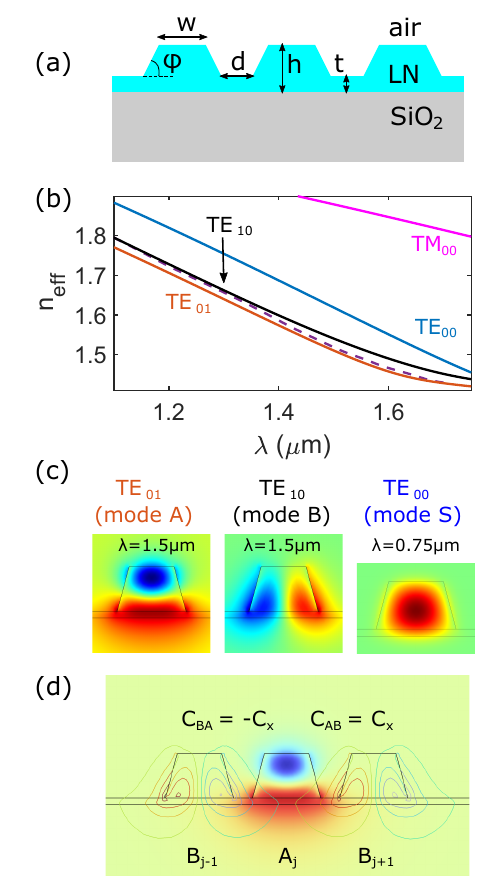}
\caption{The geometry: (a) a schematic view of an array. The extraordinary crystal axis of lithium niobate is oriented along the horizontal $x$-axis (x-cut film); (b) Dispersion of different guided modes of a single waveguide with $w=700\textrm{nm}$, $h=800\textrm{nm}$, $t=100\textrm{nm}$, $\varphi=75^\circ$. The dashed line illustrates dispersion of the edge mode in an array of $N=10$ waveguides, edge-to-edge separations $s=100\textrm{nm}$; (c)  profiles of $\textrm{TE}_{01}$ and $\textrm{TE}_{10}$ modes at $\lambda_F=1.5\mu\textrm{m}$, and $\textrm{TE}_{00}$ mode at $\lambda_S=0.75\mu\textrm{m}$ (the dominant electric field component $E_x$ is plotted); (d) modal overlaps entering the calculations of the inter-modal coupling constants in Eq.~(\ref{eq:overlap}) for $d=300$nm. Data are obtained from COMSOL Multyphisics taking into account material dispersion of lithium niobate \cite{Cai2018} and silica glass substrate \cite{agrawal}.}
\label{fig:1}
\end{figure}

The geometry considered in this work is a 1D equidistant array of identical waveguides made from a $800$nm-thick x-cut lithium niobate film on silica glass substrate, see Fig.~\ref{fig:1}(a). The top width of each waveguide is $700$nm, the sidewall angle is set to $\varphi=75^o$ (this angle is controlled by the etching process \cite{Zhu2021}), and the residual film thickness is $t=100$nm. The dispersion of different guided modes of an isolated waveguide is shown in Fig.~\ref{fig:1}(b). For the selected geometry, $\textrm{TE}_{01}$ (hereafter mode $A$) and $\textrm{TE}_{10}$ (hereafter mode $B$) have similar effective refractive indices within a wide spectral window. It is the interplay between these two families of modes which is responsible for the emergence of a non-trivial topology in the system \cite{Beer2023}. Below we will consider $\chi_2$-induced interactions between these two modes at the fundamental harmonic $\lambda_F=1.5\mu$m and $\textrm{TE}_{00}$ mode (hereafter mode $S$) at the second harmonic $\lambda_S=\lambda_F/2=0.75\mu$m. The profiles of all three modes are shown in Fig.~\ref{fig:1}(c). It is assumed that the efficient interaction between the mode $S$ at the second harmonic and the modes $A$ and $B$ at the fundamental harmonic is facilitated by a suitable quasi-phase-matching \cite{Poberaj2012}. Interactions between any other pairs of modes can be safely disregarded due to large phase miss-matches. Hence the total electric field in the $j$-th waveguide is assumed in the form:
\begin{equation}
    \vec{\mathcal{E}}_j=\left(A_j\vec{\mathbf{e}}_a+B_j\vec{\mathbf{e}}_b\right)e^{in_+\zeta-i\omega_0t}+
    S_j\vec{\mathbf{e}}_se^{i2n_s\zeta-i2\omega_0t}+c.c.\;,
    \label{eq:field_ansatz}
\end{equation}
where $j=1,2,\dots, N$, $\vec{\mathbf{e}}_{a,b,s}(x,y)$ are the spatial profiles of the three modes, $n_a=1.51$, $n_b=1.54$, and $n_s=2.10$ are the effective refractive indices of the modes, $n_{\pm}=(n_b\pm n_a)/2$, $\omega_0=k_0 c=2\pi c/\lambda_F$ is the frequency of the fundamental harmonic, and $\zeta=k_0 z$ is the dimensionless propagation distance along the waveguides. We assume that the waveguide modes are normalized to give unit power:
\begin{gather}
\left<\psi_m,\psi_m\right>=1\;,
\label{eq:inner_product}
\\
\left<\psi_a,\psi_b\right>=
     \frac14\iint\left[\vec{e}_a\times \vec{h}^*_b+\vec{e}^*_b\times \vec{h}_a\right]\hat{z}dxdy\;,
\end{gather}
where $\vec{e}_m(x,y)$ and $\vec{h}_m(x,y)$ are the electric and magnetic fields of each mode $\psi_m$, $m=a,b,s$, $\hat{z}$ is the unit vector along $z$, and the integration is performed over the cross-section of an isolated waveguide. With this normalization, $|A_j|^2$, $|B_j|^2$ and $|S_j|^2$ give power of the respective modes in the $j$-th waveguide.

\subsection{Topological properties of the array at the fundamental frequency}

First, we discuss linear properties of the array at the fundamental frequency, described by the following tight-binding model \cite{Beer2023}:
\begin{gather}
-i\partial_\zeta \begin{bmatrix}A_j\\B_j\end{bmatrix}=\hat{\Delta}\begin{bmatrix}A_j\\B_j\end{bmatrix}+\hat{C}_+\begin{bmatrix}A_{j+1}\\B_{j+1}\end{bmatrix}+\hat{C}_-\begin{bmatrix}A_{j-1}\\B_{j-1}\end{bmatrix}\;,
\label{eq:FH_lin}\\
\hat{\Delta}=\begin{bmatrix}-n_-&0\\0&n_-\end{bmatrix}\;,\;
\hat{C}_{\pm}=\begin{bmatrix}C_a&\pm C_x\\\mp C_x&C_b\end{bmatrix}\;,
\label{eq:FH_lin_matrices}
\end{gather}
where $C_{a,b,s}$ are the intra-modal coupling coefficients, $C_x$ is the coupling coefficient between the mode $A$ in waveguide $j$ and the mode $B$ in waveguide $(j+1)$. The dimensionless coupling coefficients are obtained by evaluating the overlap integrals of the modes of an isolated waveguide \cite{liu_2016}:
\begin{equation}
    \label{eq:overlap}
    C_{lm}=\frac{\omega}{k_0} \iint_{-\infty}^{+\infty} \vec{e}^*_l(x,y)\cdot\left[\Delta\epsilon\vdots\vec{e}_m(x+T,y)\right]dxdy\;,
\end{equation}
with $l$ and $m$ each being either mode $A$ or mode $B$, $T=w+d+2(h-t)\sin\varphi$ is the centre-to-centre distance between the two waveguides, $\Delta\epsilon(x,y)$ is the difference between the permittivity tensors of the two-waveguide structure (waveguides $j$ and $j+1$) and a single waveguide structure (waveguide $j$ only). Specifically, different coupling coefficients that enter the model in Eqs.~(\ref{eq:FH_lin},\ref{eq:FH_lin_matrices}) are calculated by setting $\omega=\omega_0$ and using the following combinations of modes: $C_a$ corresponds to $l=A,m=A$, $C_b$ corresponds to $l=B,m=B$, and $C_x$ corresponds to $l=A,m=B$. The overlap integrals were calculated with the help of Maxwell solver COMSOL Multiphysics.

An important feature of the above model is the variable sign of the inter-modal coupling coefficient. Due to the different parity of the two modes with respect to $x\to-x$, the integrand in Eq.~(\ref{eq:overlap}) flips the sign when calculating $C_{AB}=C_x$ (i.e. coupling between the mode $A$ in waveguide $j$ and the mode $B$ in waveguide $j+1$) and $C_{BA}=-C_x$ (i.e. coupling between the mode $B$ in waveguide $j$ and the mode $A$ in waveguide $j+1$). The corresponding modal overlaps are illustrated in Fig.~\ref{fig:1}(d). This variation of signs induces an effective chirality in the array, but it still preserves the Hermitian structure of the model \cite{Beer2023}.

\begin{figure*}
    \includegraphics[width=0.7\textwidth]{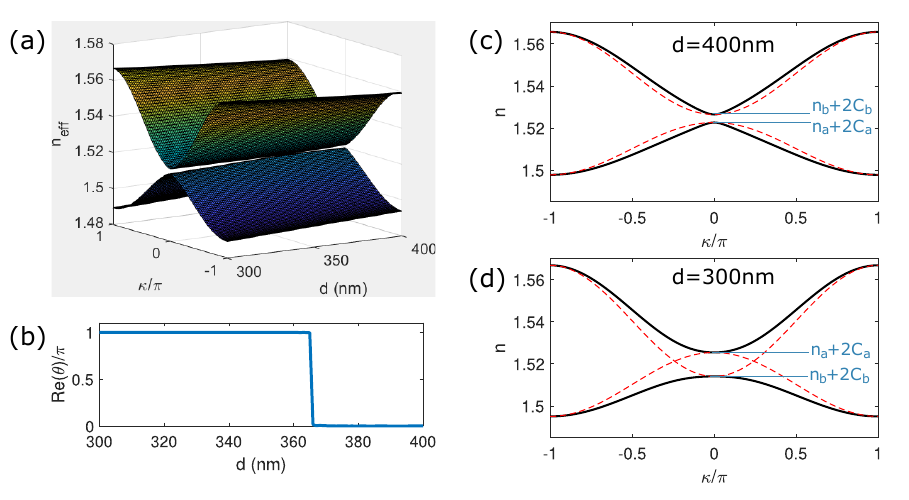}
    \caption{Spectrum of an infinite array at the fundamental harmonic: (a) two bands of the periodic structure (in a vicinity of the $\textrm{TE}_{01}$ and $\textrm{TE}_{10}$ modes of an isolated waveguide) calculated by COMSOL Multiphysics; (b) Zak phase as a function of the separation distance evaluated for the top band, a similar result is observed for the bottom band; (c) and (d) the spectrum of the tight-binding model, Eq. (\ref{eq:spectr_discr}) (solid lines) for separation distances $d=400$nm and $d=300$nm, respectively. Dashed lines indicate spectra for the case of no inter-modal coupling, $C_x=0$,  Eqs.~(\ref{eq:na_intra}), (\ref{eq:nb_intra}). For each separation, the coupling coefficients were obtained by evaluating the overlap integrals in Eq.~(\ref{eq:overlap}) using COMSOL Multiphysics.}
    \label{fig:FH_inf_spect}
\end{figure*}

The spectrum of plane waves in an infinite array, $A_j,B_j\sim \exp(in \zeta + i\kappa j)$, generally consists of two bands:
\begin{eqnarray}
    \nonumber
    n_{1,2}&=&2C_+\cos(\kappa)\\
    \label{eq:spectr_discr}
    &&\pm\sqrt{(n_-+2C_-\cos(\kappa))^2+4C_x^2 \sin^2 \kappa}\;.
\end{eqnarray}
Notably, the gap between the two bands closes either: (i) at $\kappa=0$ when $n_-+2C_-=0$; or (ii) at $\kappa=\pi$ when $n_--2C_-=0$. In our geometry $n_-=(n_b-n_a)/2$ is always positive, cf. Figs.~\ref{fig:1}(b) and (c). At the same time, $C_-$ is found to be negative within a wide range of wall-to-wall separation distances $d\in[0\textrm{nm},500\textrm{nm}]$. By varying $d$, and therefore varying $C_-$ whilst keeping $n_-$ fixed, it is possible to satisfy the condition (i). This prediction of the tight-binding model is confirmed by simulating the spectrum of an infinite periodic structure with the Maxwell solver COMSOL Multiphysics, see Fig.~\ref{fig:FH_inf_spect}(a). In our geometry, we observe that the gap closes at $d\approx 365\textrm{nm}$ and re-opens again on either side from this critical separation. This is accompanied by a topological phase transition, as can be confirmed by calculating the Zak phase of each of the band \cite{Beer2023}. For that, we evaluate modes of the periodic structure at $M=100$ discrete points across the full Brillouin zone, $\kappa_j=\pi(2(j-1)-M)/M$, $j=1,2,\dots, M$, and use the Wilson loop approach \cite{Wang2019} to calculate the phase:
\begin{equation}
\label{eq:zak}
     \theta\approx i\ln\Pi_{j=1}^{M}\left<\psi(\kappa_j),\psi(\kappa_{j+1})\right>\;,
\end{equation}
where $\psi(\kappa_j)$ are the normalized eigen-modes belonging to a particular band, the inner product is defined in Eq.~(\ref{eq:inner_product}), and $\kappa_{M+1}=-\pi$. For both bands we observe a switching between $\theta=0$ (trivial phase) to $\theta=\pi$ (non-trivial phase) at the critical separataion, see Fig.~\ref{fig:FH_inf_spect}(b).

Before and after the transition, the edges of the gap, as defined by Eq.~(\ref{eq:spectr_discr}), are given by: 
\begin{equation}
    n_{a}^{(e)}=-n_-+2C_a\;, \qquad n_{b}^{(e)}=n_-+2C_b\;.
    \label{eq:gap_edges}
\end{equation}
Note that, in accordance with the ansatz in Eq.~(\ref{eq:field_ansatz}), the spectrum of the tight-binding model is shifted by $n_+$. Hence the effective refractive indices at the edges of the gap are given by:
\begin{equation}
    n_{a}^{(e)}+n_+=n_a+2C_a\;, \qquad n_{b}^{(e)}+n_+=n_b+2C_b\;.
\end{equation}
If we neglect the inter-modal coupling, $C_x=0$, these two values correspond to the highest and lowest indices of the eigenstates formed by the intra-modal couplings of modes $A$ and $B$ respectively: 
\begin{gather}
n_{1,C_x=0}=n_a+2C_a \cos(\kappa)\;,
\label{eq:na_intra}
\\ 
n_{2,C_x=0}=n_b+2C_b \cos(\kappa)\;.
\label{eq:nb_intra}
\end{gather}
(In our geometry $C_b<0$).

The topological phase transition is accompanied by the qualitative re-structuring of the eigenmodes of the two bands \cite{Beer2023}. For separation distances larger than the critical one, $n_--2C_->0$ and therefore $n_b^{(e)}>n_a^{(e)}$. In this case, each band can be associated with a particular mode of the isolated waveguide, $A$ or $B$, and the gap is formed by the detuning between the two modes, see Fig.~\ref{fig:FH_inf_spect}(c). In contrast, for smaller separation distances the top and bottom edges of the gap are flipped, $n_b^{(e)}<n_a^{(e)}$, such that without the inter-model coupling, the two bands would be overlapping, see Fig.~\ref{fig:FH_inf_spect}(d). In this regime, the gap is formed by the inter-modal coupling, and the eigenmodes of each band are generally composed of a mixture of modes $A$ and $B$ \cite{Beer2023}.

\begin{figure*}
    \includegraphics[width=0.7\textwidth]{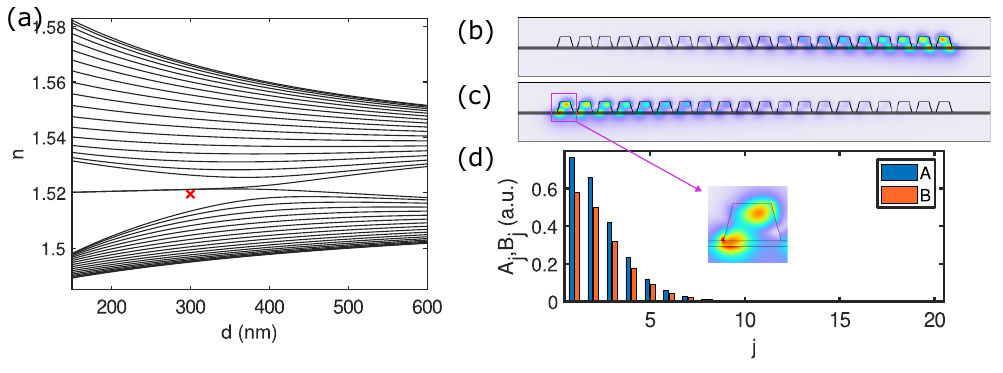}
    \caption{Edge states in $N=20$ waveguide array: (a) full spectrum of the tight-binding model as a function of the separation distance. Topological edge states emerge inside the gap at $d\le 365$nm; (b) and (c) left- and right-edge localized modes obtained by COMSOL for $d=300$nm. The corresponding effective indices (nearly degenerate) are indicated by the red crosses in panel (a); (d) the corresponding edge state as predicted by the tight-binding model.}
    \label{fig:lin_edge_states}
\end{figure*}

In accordance to the bulk-boundary correspondence \cite{Hatsugai1993, Xiao2014}, the non-trivial topology of the structure leads to the appearance of topological edge states in finite-size arrays at separation distances below the critical one. In Fig.~\ref{fig:lin_edge_states}(a) the spectrum of a $N=20$ array is plotted as a function of the separation distance $d$. This spectrum was obtained from the tight-binding model in Eqs.~(\ref{eq:FH_lin}), (\ref{eq:FH_lin_matrices}) with the coupling coefficients in Eq.~(\ref{eq:overlap}) calculated with COMSOL for different separations. As the gap closes and re-opens again, two nearly-degenerate states appear inside the gap. These two states are localized at either side of the array, and the overlap of their exponentially localized tails in the bulk of the array lifts the degeneracy. We find the results of the tight-binding model to be in good agreement with the full Maxwell solver. In Fig.~\ref{fig:lin_edge_states}(b) and (c) the profiles of the two edge states at separation distance $d=300$nm are shown, obtained in COMSOL simulations. The corresponding effective indices (practically degenerate for this size of the array) are indicated by the red cross in Fig.~\ref{fig:lin_edge_states}(a). The field intensity distribution in each waveguide suggests a superposition of the modes $A$ and $B$. The profile of the corresponding edge mode obtained from the tight-binding model is shown in Fig.~\ref{fig:lin_edge_states}(d). 

\subsection{Nonlinear interaction}

We now extend the tight-binding model to include interactions between the fundamental and second harmonic fields. For the latter, a simple single-mode model with nearest-neighbours couplings $C_s$ is assumed \cite{Gorbach2024}:
\begin{eqnarray}
\nonumber
-i\partial_\zeta \begin{bmatrix}A_j\\B_j\end{bmatrix}=\hat{\Delta}\begin{bmatrix}A_j\\B_j\end{bmatrix}+\hat{C}_+\begin{bmatrix}A_{j+1}\\B_{j+1}\end{bmatrix}+\hat{C}_-\begin{bmatrix}A_{j-1}\\B_{j-1}\end{bmatrix}\\
\label{eq:model_FH}
+\begin{bmatrix}\gamma_a S_jA_j^* e^{-i\delta\zeta}\\\gamma_b S_jB_j^* e^{-i\delta\zeta}\end{bmatrix}\;,\\
\label{eq:model_SH}
-i\partial_\zeta S_j=C_s\left(S_{j+1}+S_{j-1}\right)+\frac{e^{i\delta\zeta}}{2}\left(\gamma_a A_j^2+\gamma_b B_j^2\right)\;,
\end{eqnarray}
where $\delta=2(n_+-n_s)-2\pi/(k_0L_{qpm})$ is the phase-mismatch parameter controlled by the quasi-phase matching period $L_{qpm}$, the coupling coefficient $C_s$ is obtained from Eq.
(\ref{eq:overlap}) by setting $\omega=2\omega_0$ and $l=m=s$, and $\gamma_{a,b}$ are the effective nonlinear coefficients, which take into account overlaps of the modes $A$, $B$ and $S$ \cite{Rowe2019}:
\begin{equation}
    \gamma_{m}=\frac{c\epsilon_0}{4}\iint_{-\infty}^{+\infty}\vec{e}_s^*\cdot \left[\hat{\chi}_2 \vdots \vec{e}_m \vec{e}_m\right]dxdy\;,
\end{equation}
where $m=a,b$, the integral is taken over the cross-section of a single waveguide, $\hat{\chi}_2$ is the second-order nonlinear tensor of lithium niobate \cite{Cai2018,NL_crystals_book}.

In an infinite array, the spectrum of plane waves of the second harmonic $S\sim \exp(in\zeta+i\kappa j)$, shifted by $n=2n_s$ in accordance to the ansatz in Eq.~(\ref{eq:field_ansatz}), is given by:
\begin{equation}
n=2C_s \cos\kappa\;.
\label{eq:SH_spectr}
\end{equation}

In what follows, we will focus on the geometry with the separation $d=300$nm between the waveguides, so that the fundamental harmonic supports topological edge states. For this geometry, the following values of the dimensionless coupling coefficients and nonlinear coefficients were obtained: $C_a=0.0076$, $C_b=-0.0132$, $C_x=0.0068$, $C_s=1.35\cdot 10^{-5}$, $\gamma_a=1.1\cdot 10^{-4}\;\textrm{W}^{-1/2}$, $\gamma_b=0.4\cdot 10^{-4}\;\textrm{W}^{-1/2}$.

\section{Solitons}

We consider stationary solutions in which both harmonics are localized by virtue of the nonlinear interaction (spatial solitons). For that matter, we seek stationary solutions of Eqs.~(\ref{eq:model_FH}, \ref{eq:model_SH}) in the form:
\begin{gather}
A_j(z)=a_j\exp(i(q-\delta/2)z)\;,\\
B_j(z)=b_j\exp(i(q-\delta/2)z)\;,\\
S_j(z)=s_j\exp(i2q z)\;,
\end{gather}
where amplitudes $a_j$, $b_j$, and $s_j$ solve the following set of algebraic equations:
\begin{eqnarray}
\nonumber
\left(q-\frac{\delta}{2}\right)a_j=-n_-a_j+C_a(a_{j+1}+a_{j-1})\\
\label{eq:FHA_stac}
+C_x(b_{j+1}-b_{j-1})+\gamma_a s_ja_j^*\;,\\
\nonumber
\left(q-\frac{\delta}{2}\right)b_j=n_-b_j+C_b(b_{j+1}+b_{j-1})\\
\label{eq:FHB_stac}
-C_x(a_{j+1}-a_{j-1})+\gamma_b s_jb_j^*\;,\\
\label{eq:SH_stac}
2q s_j=C_s\left(s_{j+1}+s_{j-1}\right)+\frac{1}{2}\left(\gamma_a a_j^2+\gamma_b b_j^2\right)\;,
\end{eqnarray}
and $q$ is the free parameter.  Taking into account the symmetry of the system in Eqs.~(\ref{eq:model_FH}, \ref{eq:model_SH})
\begin{equation}
\begin{gathered}
A_j\to A_j \exp(i\alpha)\;, B_j\to B_j \exp(i\alpha)\;,\\
S_j\to S_j \exp(i2\alpha)\;,
\end{gathered}
\label{eq:sym}
\end{equation}
with $\alpha$ being an arbitrary phase shift, without loss of generality, we can assume that $a_j$, $b_j$ and $s_j$ are all real amplitudes. 

For a solution of the system of Eqs.~(\ref{eq:FHA_stac}-\ref{eq:SH_stac}) to be spatially localized, we require $2q$ to be outside of the linear band of the second harmonic, Eq.~(\ref{eq:SH_spectr}),  and simultaneously $q-\delta/2$ to be outside of the two bands of the fundamental harmonic spectrum, Eq.~(\ref{eq:spectr_discr}). Specifically, we consider solutions with $q-\delta/2$ belonging to the topological gap of the fundamental harmonic, thus:
\begin{equation}
    |q|>C_s\;,\qquad n_b^{(e)}<q-\frac{\delta}{2}<n_a^{(e)}\;,
    \label{eq:nl_mode_condition}
\end{equation}
where $n_{a,b}^{(e)}$ are defined earlier in Eq.~(\ref{eq:gap_edges}).

Fixing a value of $q$, the corresponding field profiles are found by solving numerically Eqs.~(\ref{eq:FHA_stac}-\ref{eq:SH_stac}) using Newton-Raphson iterations.

It is useful to note that Eqs.~(\ref{eq:FHA_stac}-\ref{eq:SH_stac}) have two types of homogeneous solutions (stationary points) $a_j\equiv A_{0}$, $b_j\equiv B_{0}$, $s_j\equiv S_{0}$: the $B$-type solution has zero amplitude of $A$-field
\begin{gather}
\label{eq:B-type_stac_cond}
q>C_s\;, \\
S_{0}=\frac{q-\delta/2-n_b^{(e)}}{\gamma_b}>0\;,\\
\label{eq:B_type_stac_AB_fields}
A_{0}=0\;, \qquad B_0^2=\frac{4(q-C_s)}{\gamma_b}S_0\;,
\end{gather}
and the $A$-type solution has zero amplitude of $B$-field
\begin{gather}
\label{eq:A-type_stac_cond}
q<-C_s\;, \\
S_{0}=-\frac{n_a^{(e)}-q+\delta/2}{\gamma_a}<0\;,\\
A_{0}^2=-\frac{4(C_s-q)}{\gamma_a}S_0\;, \qquad B_0=0\;.
\label{eq:A_type_stac_AB_fields}
\end{gather}
Because of the generic symmetry of the system in Eq.~(\ref{eq:sym}), the solutions with positive and negative routes for $A_0$ and $B_0$ are identical (they are related via the transformation with $\alpha=\pi$).

\subsection{Bulk solitons}

\begin{figure*}
    \includegraphics[width=\textwidth]{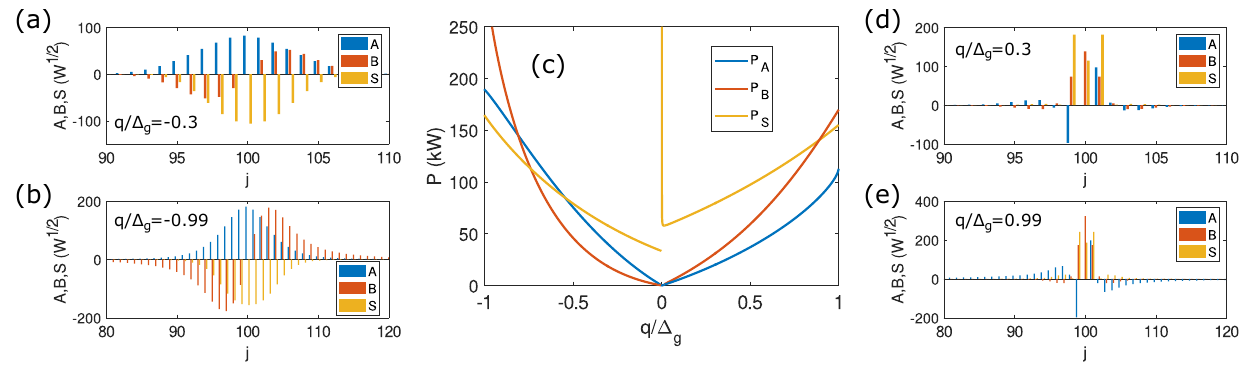}
    \caption{Bulk solitons for the phase-matching parameter value $\delta/2=-0.5(n_a^{(e)}-n_b^{(e)})$, i.e. the second harmonic field is phase matched with the mid-gap of the fundametnal harmonic. Panels (a), (b), (d) and (e) show soliton profiles for different values of the propagation constant: $q/\Delta_g=-0.3,-0.99,0.3,0.99$, respectively. Total powers of each component, $P_x=\sum_j |X|^2$, as functions of the propagation constant are plotted in panel (c).}
    \label{fig:bulk_solitons}
\end{figure*}

The two stationary points discussed above define the structure of the two types of solitons localized in the bulk of the array: each corresponding to a homoclinic orbit looping around one of the points. In Fig.~\ref{fig:bulk_solitons} both families of solitons are illustrated for the case when the phase matching is set for the mid-gap: $\delta/2=-n_0=-\left(n_a^{(e)}+n_b^{(e)}\right)/2=-(C_a+C_b)$. In this case, the $A$- and $B$-type solitons reside in the lower- and upper-half of the gap, respectively, as dictated by the conditions set in Eq.~(\ref{eq:nl_mode_condition}) and the conditions of existence of either type of the stationary point in Eqs.~(\ref{eq:B-type_stac_cond}) and (\ref{eq:A-type_stac_cond}). In Fig.~\ref{fig:bulk_solitons}(c) the total powers of each of the component of the solitons are plotted for both families as functions of the relative detuning $q/\Delta_g$, where $\Delta_g$ is the half-width of the gap:
\begin{equation}
\Delta_g=\frac{n_a^{(e)}-n_b^{(e)}}{2}=2C_--n_-\;.
\end{equation}

In $A$-type solitons, $S$- and $A$-fields have bell-shaped envelopes, typical for bright solitons, while the $B$-field has an anti-symmetric field profile with the node in the soliton core, see Figs.~\ref{fig:bulk_solitons}(a) and (b). This structure can be intuitively understood as the corresponding homoclinic orbit must cross the $B=0$ axis as it loops around the stationary point with $B_0=0$, see Eq.~(\ref{eq:A_type_stac_AB_fields}). Similar combinations of symmetric and anti-symmetric field profiles are known for two-component gap solitons in Kerr-type systems \cite{KUG1999}. The soliton becomes less localized, and the powers of all three components grow as the propagation constant $q$ shifts from the centre of the gap towards the lower edge $q=-\Delta_g$. At the lower edge, $A$- and $B$- components de-localize due to the resonance with the linear mode, see Fig.~\ref{fig:bulk_solitons}(b). At the opposite boundary, $q=-C_s$, the soliton ceases to exist due to the resonances with the linear modes of the $S$-field. Approaching $q=-C_s$, the powers of $A$- and $B$-fields tend to zero, but there is a threshold power for the $S$-component.

Similar trends are observed for the $B$-type solitons in the upper half of the gap, see Figs.~\ref{fig:bulk_solitons}(c)-(e). These solitons have an anti-symmetric shape of the $A$-field, as the homoclinic orbit loops around the $B$-type stationary point with $A_0=0$, Eq.~(\ref{eq:B_type_stac_AB_fields}). The $B$-type solitons have a more compact structure of the core, compared to $A$-type. 

\begin{figure}
    \centering
    \includegraphics[width=0.4\textwidth]{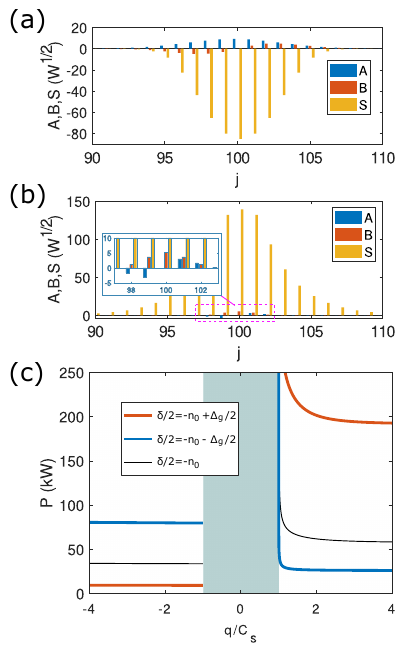}
    \caption{Bulk solitons near the boundaries $q=\pm C_s$: (a) and (b) show the structures of $A$- and $B$-type solitons at $q=-1.1C_s$ and $q=1.1C_s$, respectively, with the phase matching at mid-gap, $\delta/2=-n_0$; (c) total powers of each component as functions of the propagation constant for three values of the phase-matching parameter: $\delta/2=-n_0$ (mid-gap), $\delta/2=-n_0-\Delta_g/2$ (upper-half of the gap), and $\delta/2=-n_0+\Delta_g/2$ (lower-half of the gap). The shaded grey area indicates the forbidden region due to the constrain $|q|>C_s$ in Eq.~(\ref{eq:nl_mode_condition}).}
    \label{fig:bulk_solitons_near_midgap}
\end{figure}

\begin{figure*}
    \centering
    \includegraphics[width=\textwidth]{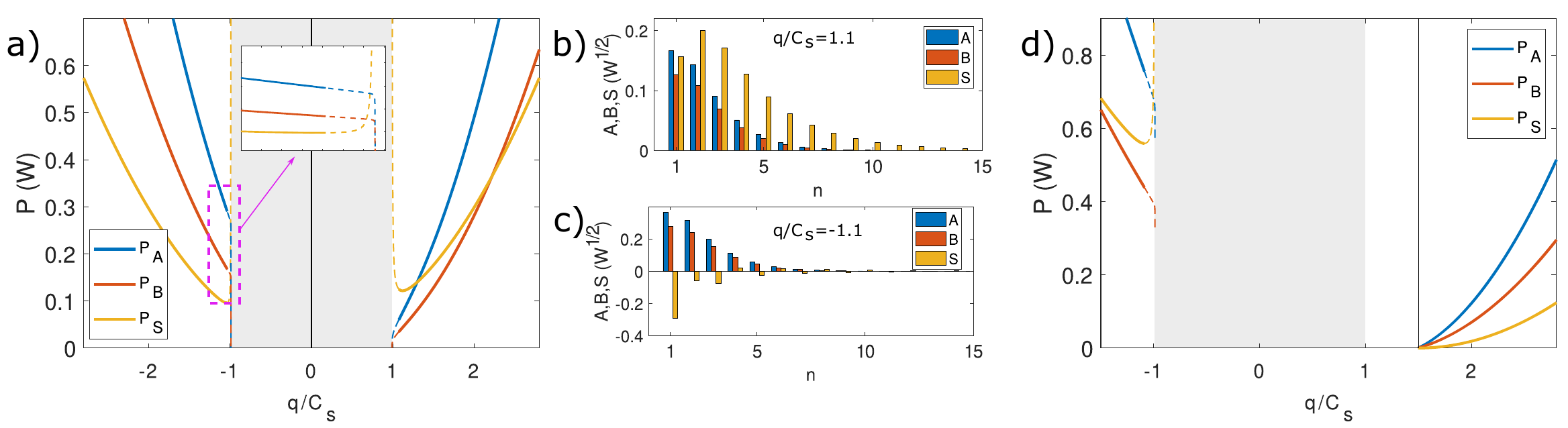}
    \caption{Edge solitons with the phase-matching parameter set to $\delta/2=-n_e$ in (a)-(c), and $\delta/2=-n_e+1.5 C_s$ in (d). In (a) and (d) the powers of the three components are plotted as functions of the mode parameter $q$, solid/dashed curves correspond to stable/unstable solutions. The shaded grey areas indicate forbidden regions due to the constrain $|q|>C_s$ in Eq.~(\ref{eq:nl_mode_condition}), and the vertical solid lines indicate the position of the linear edge mode. In (b) and (c) the soliton profiles are shown for $q/C_s=1.1$ and $q/C_s=-1.1$, respectively.}
    \label{fig:nonlinear}
\end{figure*}

The behaviour of $A$- and $B$-type solitons near the boundaries $q=\pm C_s$ is notably different. The soliton profiles near the boundaries are shown in Figs.~\ref{fig:bulk_solitons_near_midgap}(a) and (b). As the $B$-type soliton approaches the boundary $q=C_s$, its $S$-field component starts to delocalize and grow rapidly due to the resonance with the lowest wavenumber $\kappa=\pi/(N+1)$ linear mode (lowest order standing wave) of the second harmonic. This is accompanied by a rapid divergence of the power of the $S$-field soliton component, as seen in Figs.~\ref{fig:bulk_solitons}(c) and \ref{fig:bulk_solitons_near_midgap}(c). Such behaviour near the edge of a linear spectrum is typical for discrete solitons in other systems. On the contrary, the power of the $S$-field component in $A$-type soliton remains nearly constant in a vicinity of the $q=-C_s$ boundary. As seen from the plots in Figs.~\ref{fig:bulk_solitons}(a),(b) and \ref{fig:bulk_solitons_near_midgap}(a), the structure of the $S$-field component in $A$-type soliton is still similar to the lowest order standing wave, i.e. to the mode at the {\em opposite} side of the linear spectrum. Thus, the resonance with $\kappa=N\pi/(N+1)$ mode at $q=-C_s$ becomes arrested due to the different symmetry of the linear mode. It is possible to numerically continue the $A$-type soliton all the way to this boundary, where it becomes embedded in the spectrum of the linear waves of the second harmonic. However, as $q$ is shifted further inside the region $(-C_s,C_s)$, resonances with other linear modes cannot be avoided, and the soliton looses localization.

Both types of bulk solitons have power thresholds in $S$-field, as dictated by the resonances at $q=\pm C_s$ boundaries. By shifting the phase-matching either side away from the mid-gap, the threshold for one soliton type lowers, and it raises for the other type, as illustrated in Fig.~\ref{fig:bulk_solitons_near_midgap}(c). However, this tunability is limited, since the phase-matching parameter $(-\delta/2)$ must remain within the bandgap, as dictated by the condition in Eq.~(\ref{eq:nl_mode_condition}).

\subsection{Edge solitons}

Because of the non-trivial topology of the fundamental harmonic, it is also possible to find solitons localized at either edge of the array \cite{Gorbach2024}. For that, we set the phase matching near the linear topological edge mode, $\delta/2=-n_e$. In our geometry $n_e\approx 1.52$, see Fig.~\ref{fig:FH_inf_spect}(a). As in the case with bulk solitons, two different families of edge solitons (nonlinear edge modes) can be found with the propagation constant of the fundamental harmonic $(q-\delta/2)$ located on either side from the linear edge mode index $n_e$ in the topological gap, see Fig.~\ref{fig:nonlinear}. The profiles of $A$- and $B$-fields are similar in both type of edge solitons, as dictated by the profile of the linear edge mode. The main difference between the two types of solitons is in the relative phases between the fundamental harmonic and second harmonic fields, as illustrated in Figs.~\ref{fig:nonlinear}(b) and (c). The behaviour of the edge solitons in the low-power limit can be different, depending on the value of the phase-matching parameter. For $|\delta/2+n_e|<C_s$ edge solitons cannot asymptotically approach the linear edge mode limit $q_0=\delta/2+n_e$ due to the constraint $|q|>C_s$. Approaching $q=\pm C_s$, the $S$-component of the mode looses localization, its power rapidly diverges, see Fig.~\ref{fig:nonlinear}(a). As the result, in this regime edge solitons only exist above a certain power threshold. On the contrary, for $|\delta/2+n_e|>C_s$ one family of edge solitons bifurcates from the linear edge mode with zero threshold, while the power threshold for the other family still persists, as shown in Fig.~\ref{fig:nonlinear}(d). Despite the condition in Eq.~(\ref{eq:nl_mode_condition}) suggesting possible existence of nonlinear modes in the interval $1\le q/C_s\le \delta/2+n_e$, we were not able to find such modes numerically.

\section{Summary}

The recently discovered non-trivial topological phases in equidistant arrays of thin-film lithium niobate waveguides \cite{Beer2023}, emerging due to the specific interaction of different families of guided modes, make these relatively simple geometrical setups an interesting platform for studies of novel nonlinear topological phenomena. One unique feature which sets this platform apart from other topological photonic setups, based on evanescently coupled waveguides, is that the interacting fundamental and second harmonic fields experience different underlying topology. Thus, low-amplitude excitations at the fundamental frequency are described by two fields, each corresponding to a different spatial mode of an individual waveguide, interacting via inter- and intra-modal nearest-neighbour couplings. These excitations form a two-band structure with a non-trivial topology. Meanwhile, low-amplitude excitations at the second harmonic are described by a single-mode nearest-neighbour coupled oscillator type model with trivial topology. When coupled together via $\chi_2$ nonlinearity, all three components can form a localizaed stationary state, soliton, with the propagation constant of the fundamental harmonic residing in the topological gap of the two-band spectrum, and the propagation constant of the second harmonic - in the semi-infinite gap. We found that generally, there exist two families of such three-component two-colour solitons, related to the existence of two different constant-amplitude solutions (stationary points). The relative phases of the fundamental and second harmonic fields are different for the two soliton families. The propagation constant resides in the upper half of the topological gap for one family, and in the lower half - for the other.

The properties of the solitons localized in the bulk and at an edge of a waveguide array are notably different in the low-power limit. For bulk solitons, the power of the second harmonic remains finite in this limit due to resonances with linear modes. This sets a power threshold for exciting such solitons. Generally, the power thresholds for the two families of bulk solitons in the lower- and upper-halves of the topological gap are different, and can be adjusted by varying the phase-matching parameter. However, there is only a finite room for adjustments, since the phase-matching must be arranged within the finite topological gap for the solitons to exist.

On the contrary, the power thresholds for edge solitons appear to be much lower, and can be set to zero by arranging the phase-matching to be detuned from the linear edge state $n_e$ by more than one coupling strength of the second harmonic $C_s$ (but still within the topological gap). In such case, one of the two families of edge solitons appear to be bifurcating from the linear topological edge state without a threshold.

While in this work we focused on a specific geometry and setup, our earlier findings \cite{Beer2023} suggest that similar effects could be observed in such arrays within a wide range of geometrical parameters and at different wavelengths. Our discoveries pave the way to further experimental studies of topological effects in Lithium Niobate waveguide arrays.

The author declares no conflicts of interest. Data underlying the results presented are not publicly available at this time but may be obtained from the author upon reasonable request.

\bibliography{myrefs}

\end{document}